# The Possible Cause of the 40 SpaceX Starlink Satellite Losses in February 2022: Prompt Penetrating Electric Fields and the Dayside Equatorial and Midlatitude Ionospheric Convective Uplift*

**Bruce T. Tsurutani[1], James Green[2], and Rajkumar Hajra[3]**

[1]Retired, Pasadena, California USA

[2]National Aeronautics and Space Administration, Washington D.C. USA

[3]Indian Institute of Technology Indore, Simrol, Indore 453552, India..

Corresponding author: Bruce T. Tsurutani (bruce.tsurutani@gmail.com)

**Key Points:**

- Based on limited public information concerning the launch/boost/loss of Space X's Starlink satellites in February 2022, we are presenting an alternate scenario of the losses to that of Dang et al. (2022).

- The Swarm B satellite encountered enhanced mass at 500 km altitude during two moderate magnetic storms on 3 and 4 February 2022.

- The Swarm B satellite decelerations occurred primarily on the dayside at equatorial to middle latitudes.

- Our scenario is that the storm-time prompt penetrating electric fields uplifted the ionospheric $O^+$ leading to the downing of the 40 Starlink satellites.

*This paper was presented at COSPAR, Athens Greece, 16-24 July, 2022

**Abstract**

On ~1613 UT, 3 February 2022 SpaceX launched 49 Starlink satellites from Cape Kennedy, Florida into ~210 km altitude orbits where they were to receive subsequent boosts. We assume that 9 satellites received subsequent boosts and succeeded in obtaining stable orbits. Several of the Starlink satellites reentered the atmosphere in the early morning hours of 7 February. Two magnetic storms occurred one with a peak SYM-H intensity of -80 nT at ~1056 UT on 3 February and a second with a SYM-H peak intensity of -71 nT on ~2059 UT 4 February. The ESA polar orbiting Swarm B satellite data show that enhanced storm-time air mass densities occurred in dayside equatorial and midlatitudes at ~500 km altitude with a density peak increase of ~50% higher than quiet time daytime values. The nightside density peak increase was ~100-190% compared to quiet time nighttime values. Prompt penetration electric fields causing $\mathbf{E}\times\mathbf{B}$ uplift of the dayside F-region $O^+$ ions and the downward convection of the nightside $O^+$ ions can explain the Swarm B day-night asymmetry during the magnetic storm main phases and are the probable cause of the losses of the Starlink satellites. Our scenario for the 40 Starlink satellite losses is different from Dang et al. (2022) who assumed that the Starlink satellites received no boosts at 210 km altitude.

**Plain Language Summary**

Magnetic storm-time prompt penetrating electric fields (PPEFs) can cause the uplift of dayside ionospheric oxygen ions to satellite altitudes by $\mathbf{E}\times\mathbf{B}$ convection. The enhanced PPEFs during storms on 3 and 4 Febrary 2022 caused enhanced oxygen ion drag of the Space X Starlink satellites and the downing of 40 of the satellites shortly after launch.

**1 Introduction**

On 1613 UT 3 February 2022 SpaceX launched 49 Starlink satellites from Cape Kennedy, Florida. Only 9 succeeded in being boosted into a much higher stable orbit. Some of the satellites are known to have reentered the atmosphere on 7 February.

Geomagnetic storms are caused by magnetic connection between intense southwardly directed magnetic fields located within the coronal mass ejection (CME) (or the upstream "sheath" fields surrounding the CME, Tsurutani et al., 1988) and the Earth's main magnetic fields (Dungey, 1961; Gonzalez et al., 1994). The reconnected magnetic fields and solar wind plasma are convected to the midnight sector of the Earth's magnetosphere where the magnetic fields are reconnected again. The reconnected fields and plasma are jetted down the magnetotail towards the inner magnetosphere, causing auroras first in the midnight sector at absolute geomagnetic latitudes of 65° to 70° and then to absolute latitudes of 55° to 75° (the auroras occur both in the northern and southern polar regions) if the storm is particularly intense. The auroras also spread to all longitudes covering the Earth's magnetosphere at the above latitudes if the storm is intense and long lasting.

The auroras are caused by the influx of energetic ~1 to 30 keV electrons into the auroral zone atmosphere (Anderson, 1958; Carlson et al., 1998; Hosokawa et al., 2020) impacting atmospheric atoms and molecules at a height of ~110 to 90 km. The excited atoms and molecules decay emitting violet, green and red light. The influx of the energetic electrons also causes the upwelling of oxygen ions to heights where orbiting satellites will impact them causing drag on the satellites and eventual lowering of their orbits. This is the standard picture of low altitude satellite drag during magnetic storms.

Dang et al. (2022) has given a scenario for the downing of the Starlink satellites assuming that they did not receive boosts at 210 km. We will present a scenario that assumes that the satellites did receive boosts but prompt penetration electric fields have led to enhanced dayside low latitude satellite drag sufficient to cause the Starlink downing several days later.

## 2 Results

2.1 Starlink Satellite Reentry

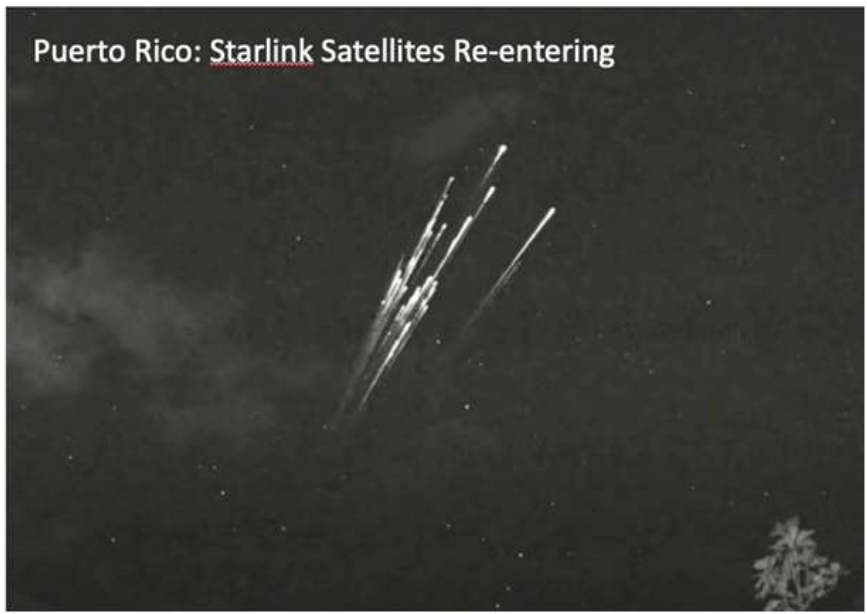

**Figure 1**. A number of the Starlink satellites reentering the atmosphere over Afiasco, Puerto Rico at ~0140 AM eastern standard time on 7 February 2022. The photograph is courtesy of Clive Simpson (https://room.eu.com/news/spacex-starlink-satellites-fall-out-of-sky-after-storm). Credit: Sociedad de Astronomia del Caribe.

Figure 1 shows that some of the Starlink satellites reentered the atmosphere in the early morning of 7 February. The satellites did not come down immediately but almost four days afterwards. The following will explain why the satellites experienced unexpected drag throughout their orbits (orbital inclination within ±51° latitude). The satellite trajectories did not go through the auroral regions (65° to 70° latitudes in both the north and the south hemisphere), and the loss was somewhat surprising.

2.2 The 3 and 4 February 2022 Magnetic Storms

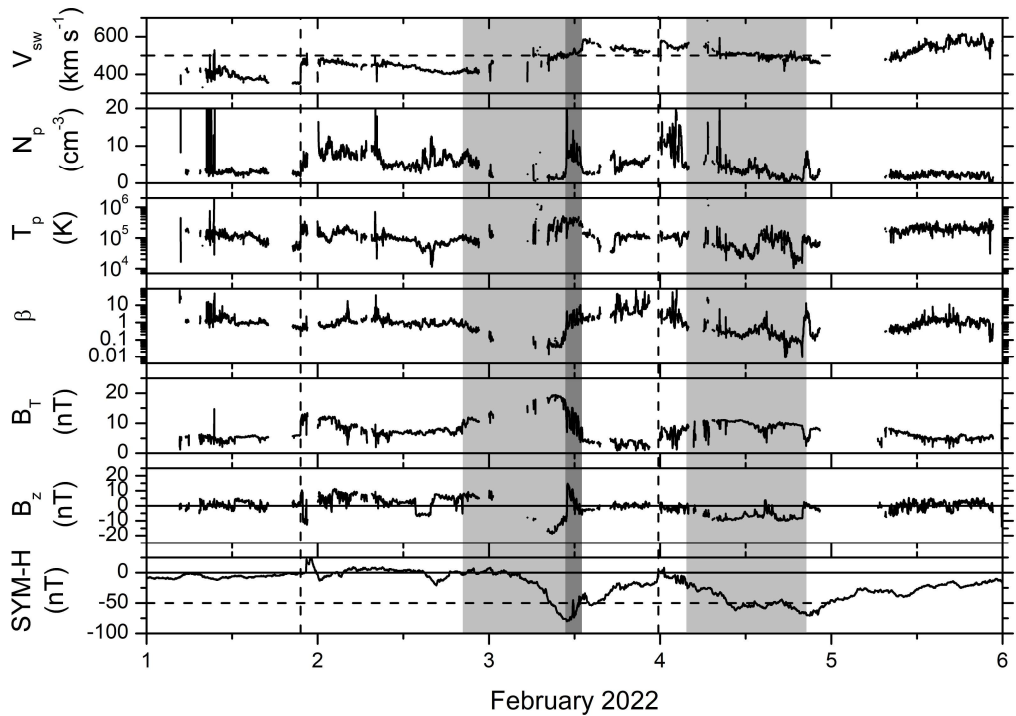

**Figure 2**. The interplanetary and geomagnetic conditions during 1–5 February 2022. The interplanetary data at 1 AU are time shifted from the spacecraft location at the L1 libration point ~0.01 AU upstream of the Earth to the nose of the Earth's bow shock. From top to bottom are: the solar wind speed ($V_{sw}$), the plasma density ($N_p$), temperature ($T_p$), plasma-$\beta$ (the magnetic pressure divided by the plasma thermal pressure), the interplanetary magnetic field magnitude ($B_T$), and the $B_z$ component. The bottom panel is the storm-time SYM-H index. The magnetic field component is given in the geocentric solar magnetospheric (GSM) coordinate system. The interplanetary data were obtained from the Goddard Space Data System (https://omniweb.gsfc.nasa.gov/) and the storm-time SYM-H index from the Kyoto University data system (https://wdc.kugi.kyoto-u.ac.jp/). The vertical dashed lines indicate interplanetary shocks. The light gray shadings

indicate magnetic clouds (MCs), and the dark gray shading indicates a solar filament propagated to 1 AU.

In Figure 2 the vertical dashed line at ~2134 UT on 1 February indicates a fast forward shock arrival. The shock causes a sudden impulse (SI$^+$) of 22 nT noted in the SYM-H index. The high-density sheath is present from the shock to the magnetic cloud (MC) portion of the ICME. The sheath did not contain major negative B$_z$ fields so was generally not geoeffective. The magnetic cloud (MC) portion of the interplanetary coronal mass ejection (ICME) is identified by high magnetic field magnitudes and low plasma-β (Burlaga et al., 1981; Tsurutani et al., 1988) and is shown by light gray shading. The MC extends from ~2100 UT on 2 February to ~1311 UT on 3 February. The interplanetary magnetic field (IMF) B$_z$ component of the MC has the characteristic "fluxrope" configuration with the magnetic field first with a negative (southward) B$_z$ component and then later a positive (northward) B$_z$ component. When the IMF B$_z$ is southward, the symmetric ring current index SYM-H decreases to a peak value of -80 nT at ~1056 UT on 3 February. Thus, the magnetic storm is caused by magnetic reconnection between the southward interplanetary fields and the Earth's magnetopause magnetic fields as postulated by Dungey (1961). This modest SYM-H peak negative value indicates a moderate intensity magnetic storm (Gonzalez et al., 1994; Echer et al., 2008). The dark gray shaded region is the high-density solar filament portion of the ICME (Illing & Hundhausen, 1986; Burlaga et al., 1998). The impact of the filament causes a compression of the magnetosphere and a sudden increase in the SYM-H index to -39 nT.

The speed of the CME at 1 AU was approximately 500 km s$^{-1}$. This is classified as a moderately fast CME (faster than the slow solar wind speed of ~350 to 400 km s$^{-1}$), thus the formation of the upstream shock and sheath. From this figure it is clear that SpaceX launched their Starlink satellites into a moderate intensity magnetic storm.

There is a second fast CME which occurred one day later. The shock is identified by a vertical dashed line at ~0004 UT on 4 February. The shock caused a SI$^+$ of intensity ~17 nT. The sheath of the second event did not contain major negative B$_z$ fields so again was generally not geoeffective. The MC portion of the second ICME is indicated by vertical shading. The MC extends from ~0225 UT to ~2012 UT on 4 February, and has a peak magnetic magnitude field of

11 nT at ~1059 UT. The $B_z$ component profile of the MC is different from the previous MC in that $B_z$ is negative or zero throughout the MC. The negative $B_z$ causes a second magnetic storm of peak intensity -71 nT at ~2059 UT on 4 February. There was no solar filament during this second ICME event.

2.3 Swarm Satellite Drag Measurements

The Swarm mission is composed of 3 satellites in polar orbit around the Earth (Olsen et al., 2013). Two of the satellites (A and C) are in circular orbit currently at an altitude of ~430 km with an inclination of ~87.4°. Swarm B is also in a circular orbit at an altitude of ~500 km with an inclination of ~88.0°. We will use only the Swarm B information here since information from Swarm A and C are redundant. Swarm A and C data show essentially the same features as will be shown in the Swarm B data. Swarm B provides data of in situ air mass density calculated from non-gravitational accelerations. The accelerations are derived from high precision orbital information from Global Positioning System (GPS) observations (van den Ijssel et al., 2020). If the satellite is impacted by any obstacle in its orbital path, e.g., increased air density, small decelerations will take place which is imprinted in its orbital evolution. Using the cross-section of the spacecraft and assumptions of the momentum change with each impact, an impact mass is calculated. The impact by air mass density is shown below.

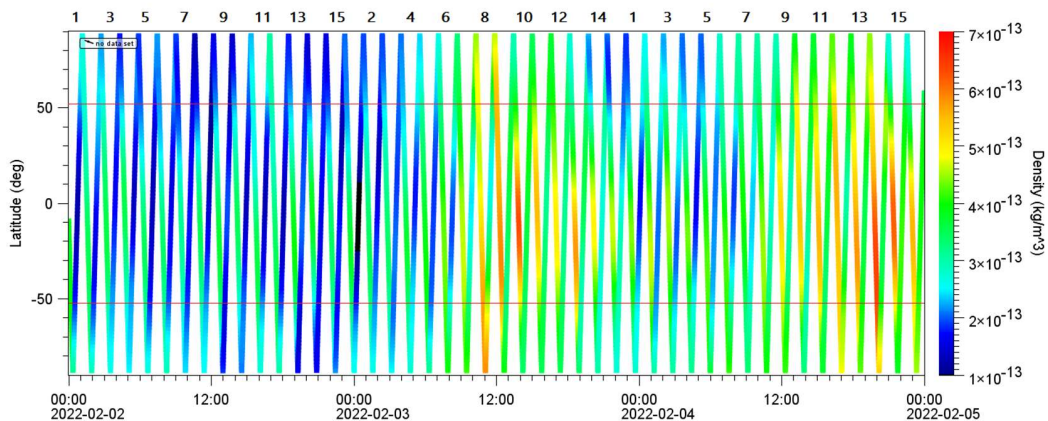

**Figure 3**. The Swarm B mass impact data for 2, 3 and 4 February 2022. The mass density is shown as a function of UT (x-axis) and geographic latitude (y-axis). It can be noted that the observations cover both day (north-to-south hemispheric passes) and night

(south-to-north hemispheric passes) sides of the globe. 2 February was a quiet day before the two magnetic storms and is shown for contrast. The two magnetic storms occurred on 3 and 4 February (Figure 2). The linear color scale is given on the right. Two red horizontal lines have been drawn in at +51° and -51° (the upper limits of the Starlink satellite orbits). Swarm B orbits on each day from the north pole to the south pole and back are marked by numbers from 1 to 15 for each day. Partial orbit 1 for 2 February is shown at the beginning of Figure 3.

Swarm B has an orbital period of ~90 minutes so there are about 15 orbits per day. To illustrate the enhanced drag, the orbits have been numbered for each day. In Figure 3 at 00 UT 2 February, the satellite is at ~ -10° latitude at ~09 local time (LT) on the dayside and is moving towards the south pole. The mass impact is a light blue color or ~$3.5\times10^{-13}$ kg m$^{-3}$. Continuing in time as the orbit crosses over the south pole and enters the nightside ionosphere at ~20 LT it is noticed that between -51° and +51° the mass impact is a dark blue color or ~$1.5\times10^{-13}$ kg m$^{-3}$. The other orbits throughout 2 February show a similar pattern between the nightside pass and the dayside pass.

On orbit 8 of 3 February, it is noted in Figure 3 that there is the first sign of a change in the mass impact at middle and low latitudes (an orange color or ~$5.0\times10^{-13}$ kg m$^{-3}$). This occurs at the south pole crossing. This happens at ~10 UT, just before the peak of the first magnetic storm. There is orange coloration throughout this downward dayside pass, across the magnetic equator and to the south pole. There is a local maximum of impact density at ~14 UT and ~09 LT at 10° latitude.

On orbits 9–13 of 3 February, the predominant density enhancements are on the dayside passes in the equatorial and midlatitude ranges. The enhancements are larger than those at higher latitudes. The maximum impact density occurred at ~19 UT and ~09 LT with a value of ~$5.5\times10^{-13}$ kg m$^{-3}$. This mass density peak increase is a factor of ~50% relative to the quiet day (2 February) daytime values.

On orbits 9–13, 3 February, the nightside equatorial and midlatitude densities are ~$3.5\times10^{-13}$ kg m$^{-3}$ indicated by the light green color. This is higher than the 2 February (quiet time) nightside densities of ~$1.5\times10^{-13}$ kg m$^{-3}$. Thus, during the magnetic storm, the nightside peak densities increased by a factor of ~100%. It is noted that the nighttime peak densities are less than the daytime peak densities. This latter feature will be explained in the Discussion section.

The high impact mass (orange color) fades out by the end of the day and doesn't start again until orbit 8 of 4 February. The light orange color (~$5.3\times10^{-13}$ kg m$^{-3}$) at the equatorial pass region on orbit 8 occurred at ~1200 UT. The slowly developing second magnetic storm started at ~ 0015 UT on 4 February, so this is approximately 10 hours after this small storm began. From orbits 8 to 11 the predominant impact mass occurs at the equatorial to middle latitudes with little or no enhanced impact mass in the auroral regions or poles. The maximum mass density occurred on dayside pass 14 with a density of ~$6.3\times10^{-13}$ kg m$^{-3}$ and extended from ~-15° to -60° latitudes. This occurred at ~2000 UT. This is coincident with the peak in the second magnetic storm. On passes 15 and 16, the impact mass decreases and the enhanced impact mass occurs mainly at the equator and middle latitudes. The maximum mass density during this second Swarm event was ~80% higher than the dayside values detected on 2 February.

The nightside pass density on orbit 14 on 4 February was ~$4.3\times10^{-13}$ kg m$^{-3}$. This is ~190% higher than comparable values on the quiet day 2 February. The nighttime peak densities are lower than the daytime peak densities, similar to the first storm features.

The data for 5 and 6 February look similar to the quiet day interval of 2 February so are not shown to conserve space.

**3 Discussion**

3.1 Magnetic Storm Prompt Penetrating Electric Fields

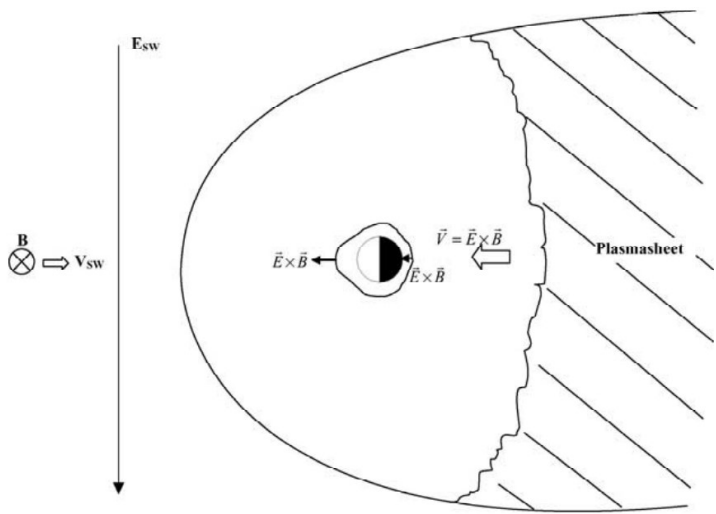

**Figure 4**. Schematic of prompt penetrating electric fields in the dayside and nightside equatorial ionospheres during magnetic storm main phases. The view is from the north pole of the Earth and the Sun is off the page on the left. The solar wind is flowing from left to right. A southward interplanetary magnetic field can be represented by a dawn-to-dusk motional electric field as shown on the left of the figure. The magnetotail magnetic reconnection and jetting of plasma into the nightside inner magnetosphere forming the storm-time ring current can be represented as a storm-time dawn-to-dusk (***E***×***B***)

convection field. The storm-time dawn-to-dusk electric field enters the equatorial and midlatitude ionosphere both in the midnight sector and also the noon sector. In the nightside ionospheric sector the electric field convects the ionosphere Earthward leading to ion-electron recombination and a decrease in total electron content (TEC) on the nightside. On the dayside, ***E×B*** convection lifts the ionospheric electrons and ions to higher altitudes. Photoionization by solar UV and EUV radiation replaces the depleted lower altitude ions giving an overall increase in TEC on the dayside. The figure is taken from Tsurutani et al. (2004) and was reproduced in Tsurutani et al. (2008).

Figure 4 was derived from a study of the 5–6 November 2001 magnetic storm using CHAMP and SAC-C GPS receiver data and TOPEX/Poseidon satellite altimeter data. GPS ground-based receivers as well as Brazilian digisonde and Pacific sector magnetometer data were used as well. The magnetic storm had a peak intensity of Dst = -275 nT, considerably higher than the two storms that occurred during 3 and 4 February 2022. This figure explained all the features of the observations for the 2001 magnetic storm.

Figure 5 is a verification of the scenario shown in Figure 4 for the 30 October 2003 magnetic storm which had a peak intensity of Dst = -390 nT (Mannucci et al., 2005). The blue trace shows the dayside TEC above the CHAMP satellite (orbiting at ~400 km altitude) before the magnetic storm onset. The two TEC peaks at ±10° geomagnetic latitude are persistent daytime ionospheric structures known as the equatorial ionization anomalies (EIAs). The red trace shows the next CHAMP daytime pass. It took place after the storm onset. The EIA peaks are located at ~ ±20° with a peak value of ~200 TEC units (a TECU is $10^{16}$ electrons m$^{-2}$). On the following dayside pass (the black trace), the EIAs are at ~ ±30° with the peak TEC values near 300 TECU. The TEC above CHAMP at ~ ±30° have increased by ~900% above quiet time values during this magnetic storm.

Why is there an increase in the EIA absolute magnetic latitudes (|MLATs|) as the storm progresses? The Earth's equatorial magnetic fields are horizontal only at the exact equator. At midlatitudes, the magnetic fields are tilted so that the ***E×B*** convection direction is to higher |MLAT| values. Why do the TEC values increase with time? When part of the ionosphere is convected to higher altitudes, the plasma recombination time scale at these altitudes is much

longer. Therefore, with the production of new ionospheric plasma at lower altitudes due to continuing solar photoionization, the overall TEC increases.

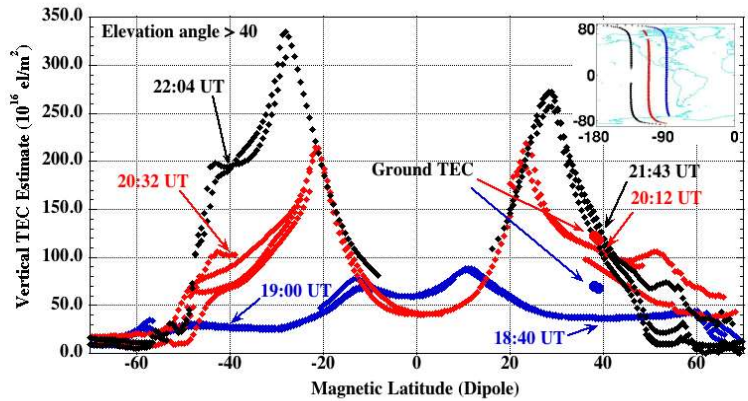

**Figure 5**. Integrated electron content (TEC) measured above the CHAMP satellite (~400 km altitude) during the geomagnetic storm of 30 October 2003. Ground tracks of the CHAMP satellite for the three dayside tracks are shown in the upper right inset. The blue curve is TEC above CHAMP prior to the storm. The red and black curves are the TEC values during the first and second passes after storm onset. The figure is taken from Mannucci et al. (2005).

The above discussion has been for TEC observations or for the column density of thermal electrons. This of course has not identified the ions which must be involved in the convective process. Following the Mannucci et al. (2005) publication, the Naval Research Laboratory (NRL) SAMI2 ionospheric code (Huba, 2000) was used to model the 30 October 2003 magnetic storm (Tsurutani et al., 2007) and its effect on ionospheric oxygen ($O^+$). The simulation showed that $O^+$ ions were convected to ~600 km altitudes and had essentially the same features as the observational ones shown in Figure 5. A quantitative examination was made and it was found that the model agreed with the Mannucci et al. (2005) CHAMP measurements within ~10%. In a subsequent work, Tsurutani et al. (2012) showed peak $O^+$ densities of $\sim 9\times 10^5$ $O^+$ ions $cm^{-3}$ at $\sim 850$ km altitude (the DMSP satellite) for the same magnetic storm.

Is there a possibility of PPEFs causing a change in the neutral oxygen densities at satellite altitudes? A possible mechanism for this is ion-neutral drag if the PPEF is intense and the upward convection of the ionosphere is rapid. Unfortunately, there is no ion-neutral drag codes available today to apply to this problem. However, a rough calculation was performed in Tsurutani et al. (2007) and they quoted "*For a 4 mV m$^{-1}$ electric field lasting for 2 h, the neutral density increase at 600 km altitude should be a factor of up to an order of magnitude greater than the quiet time values*". Lakhina & Tsurutani (2017) have performed a theoretical linear calculation for an even stronger magnetic storm, that of the Carrington 1859 event. The intensity of that storm was estimated as Dst = -1760 nT (Tsurutani et al., 2003; Lakhina & Tsurutani, 2018) with a PPEF of $\sim 20$ mV m$^{-1}$. The results of the Lakhina & Tsurutani (2017) analysis were "*It is estimated that with a prompt penetrating electric field of $\sim 20$ mV m$^{-1}$ turned on for 20 min, the O atoms and $O^+$ ions are uplifted to 850 km where they produce about 40-times-greater satellite drag per unit mass than normal.*"

Can the previous ionospheric magnetic storm results and the PPEF model (Figure 4) explain the Swarm B results presented here? The ionospheric equatorial and middle latitude dayside density increases detected by Swarm B can be explained by the PPEF model assuming that the density increases during the two magnetic storms experienced by the satellite are due to $O^+$ ions. The day-night asymmetry in Swarm B densities is what one would expect from PPEF ionospheric convection.

There were probably little or no contributions of enhanced neutral oxygen densities contributing to the Swarm B density measurements. The two magnetic storms were only moderate in intensity and the upward dayside convection of ions caused by the PPEFs would be slow. Thus ion-neutral drag can be assumed to be negligible.

## 4 Final Comments

The probable cause of the enhanced dayside drag of the Starlink satellites are the two magnetic storm main phase dawn-to-dusk directed prompt penetrating electric fields (PPEFs). The PPEFs caused ***E×B*** upward convection of the dayside F-region equatorial and middle latitude ionospheric $O^+$ ions (and equivalent number of electrons). Although the two magnetic storms were not particularly intense, the dayside density enhancements were definitely measureable by the highly sensitive Swarm B instrumentation. The dayside enhancements were peak values of ~50% on the dayside and peak values of ~100 to 190% on the nightside. These values are in good accord with Starlink statements that ~50% density increases caused the downing of their 40 satellites.

Dang et al. (2022) has provided a scenario for the downing of the Starlink satellites assuming that they had no orbital boosts at 210 km altitude. The authors mention that at that altitude "depending on the satellite cross section area, the satellites would be lost in 5 to 7 days. The storm density increase would shorten the lifetimes by about a half day" (J. Lei, private communication, 2022). Although we have not had any Starlink orbital information from Space X after months of trying, it seems inconceivable that the Starlink satellites would not have received orbital boosts. We have therefore provided an alternative scenario for the downing of the satellites several days later.

## Acknowledgments

This paper is dedicated to the memory of Y. Kamide who was a pioneer in the study of electric fields in the magnetosphere and ionosphere. The authors would like to thank Yaireska Collado-Vega, Dogacan Ozturk, Anthony Mannucci and Joe Huba for insightful discussions of the Starlink events. The Swarm mission is operated by the European Space Agency. Swarm data is publicly available at https://swarm-diss.eo.esa.int.

## Funding:
Science and Engineering Research Board Contract No. SB/S2/RJN-080/2018 (RH)

## Author contributions:
Conceptualization: BTT
Methodology: BTT

Investigation: BTT, JG, RH

Visualization: RH, BTT

Writing – original draft: BTT

Writing – review & editing: BTT, JG, RH

**Competing interests:** Authors declare that they have no competing interests.

**Data and materials availability:** The interplanetary data were obtained from the Goddard Space Data System (https://omniweb.gsfc.nasa.gov/) and the SYM-H index from the Kyoto University data system (https://wdc.kugi.kyoto-u.ac.jp/). The Swarm satellite data are obtained from https://swarm-diss.eo.esa.int.